\documentclass{aa}

\usepackage{graphicx}
\usepackage{lscape}

\newcommand{\brg}{Br$\gamma$}

\begin{document}
\title{The young Galactic star cluster [DBS2003]\,179\,\thanks{Based
on observations collected with the Very Large Telescope of the European
Southern Observatory within the Observing Program 79.D-0149(A).This paper
includes data gathered with the 6.5 meter Magellan Telescopes located
at Las Campanas Observatory, Chile.}
}
\author{J.\ Borissova\inst{1}
\and
V.D.\ Ivanov\inst{2}
\and
M. M.\ Hanson\inst{3}
\and
L.\ Georgiev\inst{4}
\and
D.\ Minniti\inst{5}
\and
R. \ Kurtev\inst{1}
\and
D.\ Geisler\inst{6}
}

\institute{Departamento de F\'isica y Astronom\'ia, Facultad
       de Ciencias, Universidad de Valpara\'{\i}so,
       Ave. Gran Breta\~na 1111, Playa Ancha, Casilla 5030,
       Valpara\'iso, Chile; \email{jura.borissova@uv.cl; radostin.kurtev@uv.cl}
         \and
European Southern Observatory, Av.\ Alonso de C\'ordoba 3107, Casilla 19,
Santiago 19001, Chile; \email{vivanov@eso.org}
\and
University of Cincinnati, Cincinnati, Ohio 45221-0011, USA; \email{ margaret.hanson@uc.edu}
       \and
  Instituto de Astronom\'ia, Universidad Nacional Aut\'onoma de M\'exico, Apartado Postal 70-254, CD
  Universitaria, CP 04510 M\'exico DF, M\'exico; \email{georgiev@astroscu.unam.mx}
   \and
Departmento de Astronom\'ia y Astrof\'isica, Pontificia Universidad Cat\'olica de Chile, Av. Vicu\~na Mackenna 4860, 782-0436 Macul, Santiago, Chile and Vatican Observatory, Specola Vaticana, V00120 Vatican City State, Italy; \email{dante@astro.puc.cl}
       \and Departmento de Astronom\'{\i}a, 
Universidad de Concepcion, Casilla 160-C, Concepcion, Chile; \email{dgeisler@astro-udec.cl}}

\offprints{J. Borissova}

\date{Received..... accepted..... }

\authorrunning{Borissova et al.}
\titlerunning{Near-infrared spectroscopy of [DBS2003]\,179}

\abstract
	{Recent near- and mid-infrared surveys have brought evidence that the Milky
Way continues to form massive clusters.}
	{We carry out a program to determine the basic physical properties
of the new massive cluster candidate [DBS2003]\,179.}
	{Medium-resolution K-band spectra and deep near-infrared images of
[DBS2003]\,179 were used to derive the spectral types of eight member stars,
and to estimate the distance and reddening to the cluster.}
	{Seven of ten stars with spectra show emission lines. Comparison
with template spectra indicated that they are early O-type stars. The
mean radial velocity of the cluster is V$_{\sc rad}$=$-$77$\pm$6 ~\,km\,s$^{-1}$.
Knowing the spectral types of the members and the color excesses, we
determined extinction $A_V$$\sim$16.6 and distance modulus
($m$$-$$M$)$_0$$\sim$14.5\,mag (D$\sim$7.9\,Kpc). The presence of early
O-stars and a lack of red supergiants suggests a cluster age of 2-5\,Myr. 
The total cluster mass is approximated to 0.7$\times$10$^4$\,$M_{\odot}$
and it is not yet dynamically relaxed.}
	{The candidate [DBS2003]\,179 further increases the family of the massive young clusters
in the Galaxy, although it appears less massive than the
prototypical starburst clusters.}


   \maketitle

%

\section{Introduction}
Why does our Galaxy contain only a few very massive young clusters?
A cluster with an initial mass of 
$10^5$\,$M_\odot$ is expected to disrupt in about 5 Gyrs, while a 
$10^6$\,$M_\odot$ cluster has a lifetime well in excess of Hubble time
(Lamers et al. 2005). According to Larsen (2006), the disk should then 
contain about 500 clusters with masses greater than $10^5$\,$M_\odot$ 
and 80 clusters with M $>$ $10^6$\,$M_\odot$ formed over its lifetime. 
Excluding the Galactic center region, only three very massive young 
clusters are known: Westerlund 1 (Clark et al. 2005), RSGC1 (Figer et 
al. 2006), and RSGC2 (Davies et al. 2007). They have masses in the range 
$10^4$-$10^5$\,$M_\odot$ and ages of $< 20 \times 10^6$ years.  
The low number of such clusters in our Galaxy was interpreted by Larsen 
(2006) either as a truncation of the cluster initial mass function (CIMF) in 
the vicinity of $10^5$\,$M_\odot$, or that the Galaxy's CIMF slope is much 
steeper than the power-law slope of $-$2 seen in other galaxies. Thus, 
the questions of the recent star formation in the disk of our Galaxy and the slope 
of the CIMF are still unanswered. Several approaches can 
be used: a detailed investigation of the CIMF (Portegies Zwart et al. 2007, 
Dowell et al. 2008, Stolte et al. 2005, etc.); an accurate 
estimation of the disruption effects (Lamers et al. 2005); 
and finally a study to find out if the cluster mass distribution is really 
independent of the star formation history (Portegies Zwart et al. 2006).  

More than 2000 infrared ``hidden'' star cluster {\it candidates} have 
been discovered in the last 5-6 years from all-sky IR surveys such as 
the 2\,MASS  (Skrutskie et al. 1997) and the {\it Spitzer Space 
Telescope} Galactic Legacy Infrared Mid-Plane Survey Extraordinaire
(GLIMPSE, Benjamin et al. 2003) and can be used to test the recent 
star formation in the Galaxy.  

This work is a part of a larger program aimed at characterizing the hidden current star population in the Galaxy (Ivanov et al. 2002, 2005; Borissova et al. 2003, 2005, 2006; Kurtev el al. 2007). Our main goal is to determine if the Milky Way still forms fairly massive star clusters at a rate similar to the rate seen in other normal spiral galaxies (Larsen 2002).

\begin{figure}[h]
\resizebox{\hsize}{!}{\includegraphics{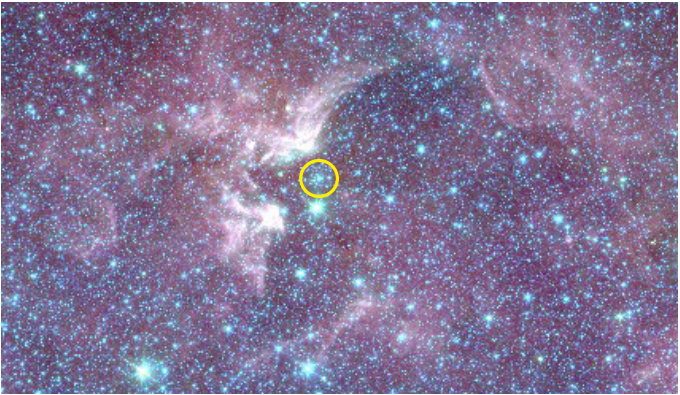}}
\caption{Pseudo-true color rendering of the region around [DBS2003]\,179 outlined with a yellow circle. 
Blue, green, and
red represent the Spitzer [3.6],[4.5], and [5.8] bands. The field of
view is 0.3$\times$1.0\, degree. North is up, and East is to the left.}
\label{db179_glimpse}
\end{figure}

The open cluster candidate [DBS2003]\,179 was discovered by Dutra et al. (2003) via visual inspection of 2\,MASS images during their survey of embedded clusters in the areas of known radio and optical nebulae. 
The cluster is located in the direction toward the H{\sc II} ionized region
G347.6+0.2 (Fig.~\ref{db179_glimpse}). Caswell \& Haynes (1987) measured
peak brightest temperature of 5\,K at 5\,Ghz continuum and angular size
of 9\,arcmin for G347.6+0.2.

Borissova et al. (2005; hearafter Paper~I) obtained high quality
infrared (IR) photometry and confirmed that [DBS2003]\,179 is a member
of the large family of ``newly'' discovered heavily obscured clusters.
Based on the photometric analysis and on the so called ``10$^{\rm th}$
brightest star method'' (Dutra et al 2003a), they determined some of
the cluster parameters: $A_V$=19, ($m$$-$$M$)$_0$=13.5\,mag, age of
$\sim$7\,Myr, and total mass of $\sim$5.5$\times$10$^3$\,$M_{\odot}$.
Thus, the cluster seems to be young and moderately massive.

However, the photometry alone is not sufficient to derive an accurate
distance to the star cluster and the uncertainty of the 10$^{\rm th}$
brightest star method can easily be as large as
$\sigma${($m$$-$$M$)$_0$}$\sim$2.5\,mag or a factor of $\sim$10 in
distance (see the discussion in Paper~I). To obtain more accurate
physical parameters for this cluster we embarked on a project
to determine the spectral types of some of the cluster members using
K-band spectroscopic observations.

The next section describes the data.  In the third section, we present
the spectral classification of the observed stars and their radial
velocities.  In Sect.~4 the cluster parameters and the total mass of
the cluster are determined. The last section is a summary of the
results.


\section{Observations and data reduction}

We carried out near-IR broad $JHK_S$ and narrow band imaging on Aug 2, 2003 with the near-IR imager Persson's Auxiliary
Nasmyth Infrared Camera (PANIC) at the 6.5-meter Baade telescope at the Las
Campanas Observatory. The instrument uses a 1024$\times$1024 HgCdTe
Hawaii detector array. The scale is 0.125\,arcsec\,$\rm pixel^{-1}$,
giving a total field of view of 2.1$\times$2.1\,arcmin. The Br$\gamma$
filter is centered on 2.165$\mu$m and has a width of 0.03\,$\mu$m.
The images are reduced following the ``standard'' procedures for the
IR, the photometry is obtained using the DAPHOT package within
Image Reduction and Analysis Facility (IRAF)\footnote{IRAF is distributed by the National
Optical Astronomy Observatories, which are operated by the Association
of Universities for Research in Astronomy, Inc., under cooperative
agreement with the National Science Foundation.}
and the typical photometric uncertainties are 0.05-0.06\,mag. We refer
the reader to Paper~I for more details. Since the goal of the
narrow-band observations was to select the sources with IR excess, no flux 
standards were taken and no attempt was made to transform the
Br$\gamma$ instrumental magnitudes to any standard system.

\begin{figure}
\includegraphics[width=\columnwidth]{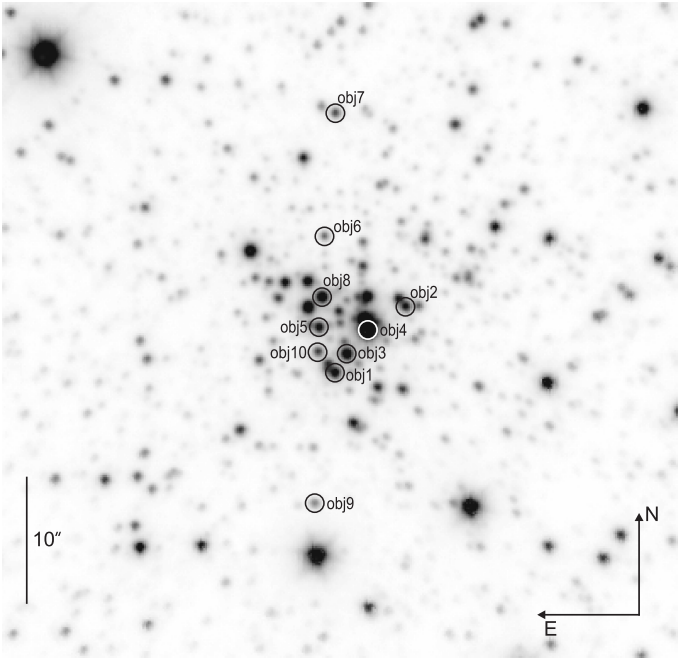}
\caption{All stars observed spectroscopically with ISAAC at the
ESO VLT are marked. The field of view is 50$\times$40\,arcsec.
}
\label{db179_chart}
\end{figure}

Based on their position in the statistically-decontaminated near-IR
color-magnitude diagram given in Paper~I, four probable members of the
[DBS2003]\,179 were selected for follow-up spectroscopy. Two of
them were among the brightest objects and the rest are chosen to be the
8th and 10th brightest cluster members. Medium-resolution ($R$=9000) K-band
spectra were obtained with Infrared Spectrometer And Array Camera (ISAAC) (Moorwood et al. 1998) on the ESO
VLT at Paranal, Chile in service mode, in May 2007. We used the
SWS1-SW mode, 0.3~arcsec wide slit, and the grating was centered at
2.134\,$\mu$m. Two radial velocity standards HD191639 (B1V,
V$_{\sc rad}$=$-$7\,km\,s$^{-1}$) and HD167785(B2V,
V$_{\sc rad}$=$-$10.6\,km\,s$^{-1}$) were also observed with an identical
spectroscopic set up. We observed with two slit positions. The exposure
times were 20\,min, the seeing during the observations was less than
1\,arcsec. However, some additional stars fell into the long slit
(120\,arcsec length).  We present spectra of ten stars in total for
the two slit positions (Fig.~\ref{db179_chart}). Their coordinates and
magnitudes are given in Table~\ref{stars_mag}.
The near-IR magnitudes are from Paper~I. The Br$\gamma$ instrumental
magnitudes are from this work. The mid-IR magnitudes are from the
Galactic Legacy Infrared Mid-Plane Survey Extraordinaire (GLIMPSE;
Benjamin et al. 2003) archive. The ``sat.'' marks a saturated star.
 
\begin{table*}\small
\begin{center}
\caption{Spectroscopically observed stars in [DBS2003]\,179.}
\label{stars_mag}
\begin{tabular}{lllllllrrrr}
\hline
& & & & & & & & & & \\[-6pt]
\multicolumn{1}{l}{Object} &
\multicolumn{1}{l}{RA(J2000)} &
\multicolumn{1}{l}{DEC(J2000)} &
\multicolumn{1}{c}{$J$} &
\multicolumn{1}{c}{$H$} &
\multicolumn{1}{c}{$K_{s}$} &
\multicolumn{1}{c}{$Br\gamma$} &
\multicolumn{1}{c}{[3.6]} &
\multicolumn{1}{c}{[4.5]} &
\multicolumn{1}{c}{[5.8]} &
\multicolumn{1}{c}{[8.0]} \\
&hh:mm:ss&deg:mm:ss & mag  & mag & mag &ins.mag &mag  &mag & mag & mag\\
& & & & & & & & & & \\[-8pt]
\hline
& & & & & & & & & & \\[-6pt]
obj1& 17:11:32.14&-39:10:49.74&14.03&12.27&11.55&15.56& 9.717& 9.378& 9.397&9.181\\
obj2& 17:11:31.73&-39:10:44.28&14.83&12.92&11.96&16.03& & & &  \\
obj3& 17:11:32.03&-39:10:48.76&13.63&11.93&11.32&15.19& & & &  \\
obj4& 17:11:31.89&-39:10:47.15&12.25&sat. &sat. &13.24& 7.657& 7.320& 7.154&7.030\\
obj5& 17:11:31.85&-39:10:50.77&14.23&12.55&11.80&15.89& & & &  \\
obj6& 17:11:31.27&-39:10:50.10&16.69&14.54&13.43&17.51& & & &  \\
obj7& 17:11:30.48&-39:10:48.94&15.62&13.78&12.90&17.14&11.942&11.766&11.598&  \\
obj8& 17:11:31.65&-39:10:50.46&14.06&11.99&11.07&14.99& & & &  \\
obj9& 17:11:33.31&-39:10:51.55&13.23&10.33&8.911&13.10&	7.991& 7.950& 7.610&7.500\\
obj10&17:11:32.00&-39:10:50.89&15.51&13.59&12.68&16.86&	& & &  \\
& & & & & & & & & & \\[-8pt]
\hline
\end{tabular}
\end{center}
\end{table*}

The spectra were dark-subtracted and flat-fielded using the ESO ISAAC
pipeline. To correct for telluric emission the object was ``nodded''
between two positions on the slit (A and B) so that the background of
position A is recorded while the source is at position B and is
subtracted from the observation when the source is at position A,
and vice-versa. The offset between the A and B positions was chosen
to avoid source overlap. The spectra were extracted using the IRAF
task {\tt apall}. For each position
along the slit a night sky spectrum was extracted with the same
extraction parameters. The wavelength calibration of the spectra are
performed with the imprinted sky emission lines. The accuracy of
the wavelength calibration is $\sim$5-7\,km\,s$^{-1}$. Telluric
absorption lines were removed using the telluric star Hip 087164 
of spectral type B2, observed under identical sky conditions and 
using the method described in Hanson el al. (1996, 2005) and Bik et 
al. (2005, 2006). First, the photospheric lines (\brg \, and HeI) in the 
spectrum of the telluric B standard was divided out. 
We did this with the template spectrum of the
B2 star HD36166 from the spectral atlas of \cite{han05}.  
The error on the resulting \brg\ equivalent width
(EW) of our target star is about 10~\%. 
After the removal of the stellar lines from 
the telluric standard star we remove the telluric lines from
the program objects by taking the ratio of the target
spectrum with this telluric spectrum. We did the telluric-line
correction using the IRAF task {\tt telluric}, which allows
for a shift in wavelength and a scaling in line strength yielding a
more accurate fit. The task uses a cross-correlation procedure to
determine the optimal shift in wavelength and the scaling factor in
line strength. The shifts are usually less than one pixel (1 pixel
corresponds to 17\,km\,s$^{-1}$).

\section{Reddening and spectral classification of the spectroscopic targets}

\subsection{Reddening}
\begin{figure}
\resizebox{\hsize}{!}{\includegraphics{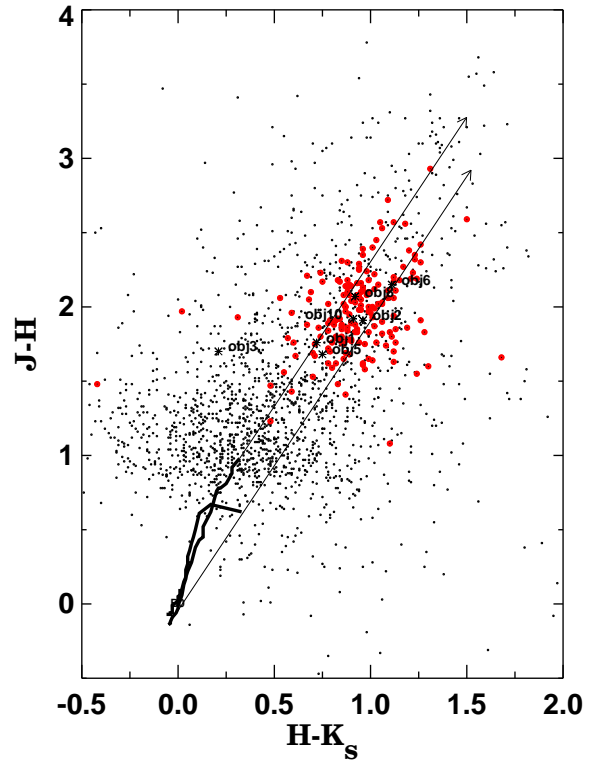}}
\caption{The $J$$-$$H$ versus $H$$-$$K_S$ color-color diagram of
[DBS2003]\,179. All stars in the field of view are shown as small
circles, while the candidate cluster members are plotted with large filled
circles. The solid lines represent the unreddened sequence of
stars of luminosity classes I and V, taken from Koornneef (1983)
and \cite{sch82}. The spectroscopically observed stars are labeled.
Reddening vectors for O5V and M5I stars are also shown.
}
\label{db179_jhk}
\end{figure}

Typically, the location of the stars in the color-color diagrams such as
 $J-H$ vs. $H$$-$$K_S$ allow us to separate between cluster members
and field red giants and to estimate their individual extinction
values. The two color diagram of all stars in our field of view is
shown in Fig.~\ref{db179_jhk}. The most probable statistically-selected 
cluster members taken from Paper~1 are plotted with large
filled circles. They are determined using a comparison field, 
with an area equal to that of the cluster.
The solid lines represent the unreddened sequence
of stars of luminosity classes I and V, taken from  Koornneef (1983)
and \cite{sch82}. The majority of the cluster members
occupy a well-defined locus in this diagram at
$H$$-$$K_S$$\approx$0.7-0.8 and $J$$-$$H$$\approx$2\,mag.

 The intrinsic $H$$-$$K_S$ and $J$$-$$H$ color of OB stars vary
between $-$0.05 and $-$0.2 (Schmidt-Kaler 1982, Koornneef 1983). 
Furthermore, the calibrations of Martins \& Plez (2006) give the same colors 
for the O3I-O9I; the O3III-O9III and the O3V-O9V type stars: 
($J$$-$$K$)$_0$=$-$0.21,
($J$$-$$H$)$_0$=$-$0.11, ($H$$-$$K$)$_0$=$-$0.1. We use these to
calculate the individual color excesses of the spectroscopically-classified stars: 
E($J$$-$$K_S$), E($J$$-$$H$) and E($H$$-$$K_S$)
are given in Table~\ref{reddening}. We calculated the E(B-V) using
the color excess ratios of E($J$$-$$K$), E($H$$-$$K_S$), and
E($J$$-$$H$) from Bessell et al. (1998) and Mathis (1990). 
The errors of E(B-V) reddening
determination  given in Col.\,6 are calculated as a standard
deviation of the mean value and quadratically added errors from
photometry and the variance due to different intrinsic estimations.
In the last column of Table~\ref{reddening} the extinction
corrected $K_S$ magnitude are given. We adopted E($B$$-$$V$)=5.2 for the mean reddening of the cluster and
thus $A_V$=16.64, using the standard reddening low.

\begin{table*}
\begin{center}
\caption{Individual reddening of the spectroscopically observed
stars in [DBS2003]\,179.}
\label{reddening}
\begin{tabular}{llllllr}
\hline
& & & & & & \\[-6pt]
\multicolumn{1}{l}{Object} &
\multicolumn{1}{l}{E($J$$-$$K_S$)} &
\multicolumn{1}{l}{E($J$$-$$H$)} &
\multicolumn{1}{l}{E($H$$-$$K_S$)} &
\multicolumn{1}{l}{E($B$$-$$V$)} &
\multicolumn{1}{l}{$\sigma$} &
\multicolumn{1}{l}{$K_0$}\\
& & & & & & \\[-8pt]
\hline
& & & & & & \\[-6pt]
obj1 &2.67&2.00&0.77&4.69&0.22& 9.90\\
obj2 &3.06&2.15&1.01&5.47&0.09&10.04\\
obj3 &2.50&1.94&0.66&4.34&0.27& 9.79\\
obj5 &2.62&1.92&0.80&4.64&0.16&10.17\\
obj8 &3.18&2.31&0.97&5.61&0.19& 9.10\\
obj10&3.02&2.16&0.96&5.27&0.13&10.79\\
& & & & & & \\[-8pt]
\hline
\end{tabular}
\end{center}
\end{table*}

\subsection{Spectral classification}

We performed a preliminary spectral classification of the program stars, using direct comparison of spectra between a star of known spectral
and luminosity class and a star which is unknown. The appropriate template spectra of early stars are taken from spectral atlas of \cite{han05}. Some of the spectra from this atlas are taken with the Subaru telescope and are of higher spectral resolution ($R$=12000) and are rebinned to $R$=9000. In general, our K-band spectra can be divided into two different classes: seven objects are emission line stars and in three of them we can identify only absorption lines.

Our spectral setting, centered at 2.134\,$\mu$m, contains the
following lines for the hot stars: Brackett\,$\gamma$ (4-7)
2.1661\,$\mu$m (Br$\gamma$); He{\sc ii} $\lambda$2.188 (7-10);
He{\sc i} $\lambda$2.1127 ($3p\ ^3$P$^o-4s\ ^3$S, triplet), and
$\lambda$2.1138 ($3p\ ^1$P$^o-4s\ ^1$S, singlet). Among the
hottest stars, the C{\sc iv} triplet ($3d^2D-3p^2P^o$) at 2.069,
2.078 and 2.083 is also seen. Hanson et al. (1996) identified the
broad emission feature found at 2.1155$\mu$m as N{\sc iii} (7-8).
Recently, Geballe et al. (2006 ) showed that O {\sc iii} 2.116 $\mu$m 
(8-7) transitions feature is an important contributor to the 2.11 $\mu$m 
emission complex.

\subsubsection{Stars showing emission lines}

\begin{figure}
\resizebox{\hsize}{!}{\includegraphics{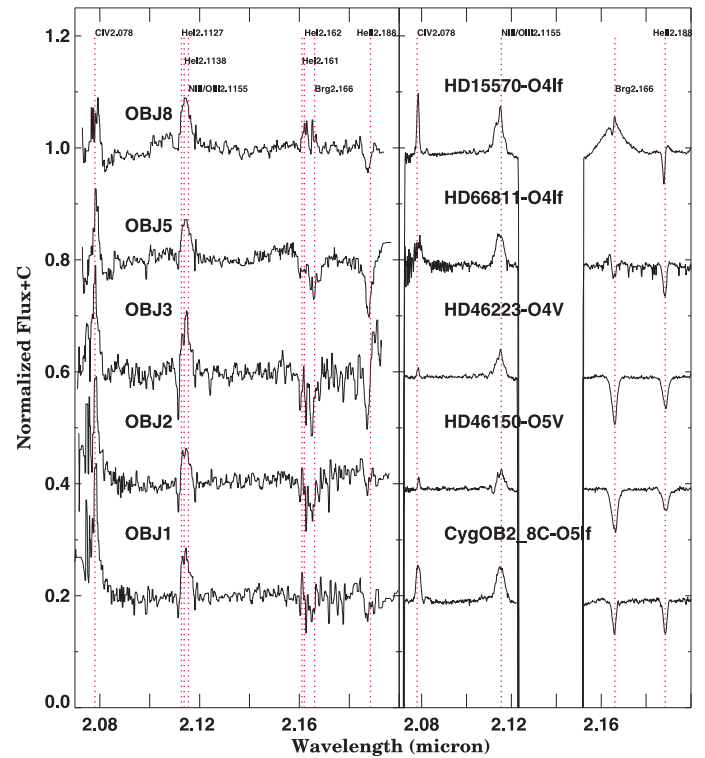}}
\caption{Left panel: The emission line stars of [DBS2003]\,179.
The different spectra have been arbitrarily shifted along the
flux $F_{\lambda}$ axis for clarity. The dashed lines indicate
the rest-frame wavelengths of the spectral features in the K-band.
Right panel: The  classification template spectra taken from \cite{han05} that 
show similarities to the cluster stars. 
}
\label{db179_sp_em}
\end{figure}

\begin{figure}
\resizebox{\hsize}{!}{\includegraphics{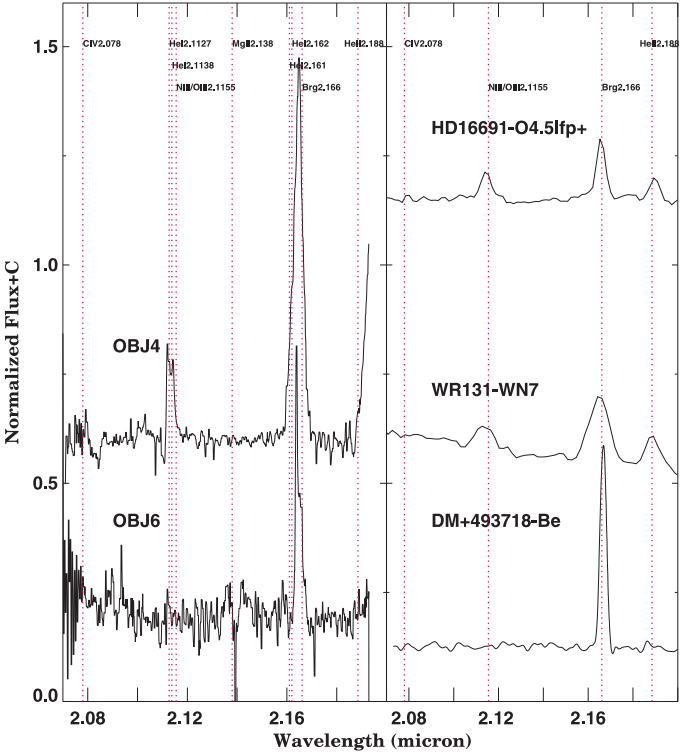}}
\caption{Left panel: The emission line stars of [DBS2003]\,179.
Right panel: The template spectra taken from \cite{han96}.
The symbols are the same as in Fig.~\ref{db179_sp_em}.
}
\label{db179_sp_em1}
\end{figure}

The spectra of the stars showing emission lines 
are shown in Figs.~\ref{db179_sp_em} 
and ~\ref{db179_sp_em1}. The Equivalent Widths (EWs) of their 
emission lines are listed
in Table~\ref{stars_spectra}. They are measured on the continuum
normalized spectra, by the IRAF task {\tt splot}, using the
deblending function. The errors of the EW are estimated taking
into account the signal-to-noise ratio (see Col. 2 of
Table~\ref{stars_spectra}), the peak over continuum ratio of the
line (see Bik et al. 2005) and the error from the telluric star
subtraction and are estimated to be $\sim$10-15\% in the worst
cases. The EW of the emission lines are negative.
The symbol ``Em'' given in
Col. 1 of the Table~\ref{stars_spectra} indicates that the line 
is in emission, ``NP'' means that the line is not present in 
the spectrum.

The C{\sc iv} line is at the limit of our setup, where the errors
are large and it is not given strong consideration. The N{\sc iii}/O{\sc iii}
emission complex appears in all these stars, except for Obj6, where it is
not present. The Br$\gamma$ line appears weakly in absorption in Obj1
(EW=0.8\AA), moderately weak in absorption in Obj2, Obj3, and Obj5 (EW=3.4-3.7\AA) 
and weakly in emission in Obj 8(EW=$-$1.4\AA). The Br$\gamma$ line appears strongly in 
emission in Obj4 and Obj6, (EW=$-$37.2\AA\, and EW=$-$12.1\AA, respectively).
The He{\sc ii} $\lambda$2.188 line is seen in absorption in Obj1, 2, 3, 5, and 8.

We derive the following spectral classification of our objects:

  Object\,1 and Object\,2 are quite similar, given the limitation
of the ``edge'' effects, where both the C{\sc iv} 2.078\,$\mu$m and
He{\sc ii} 2.1885\,$\mu$m lines are compromised. The EW of the
N{\sc iii}/O{\sc iii} line of Obj1 is $-$2.8\AA\,  which, according to
Hanson et al. (2005) defines the star as later than O3-O4, but not
later than O6, when He{\sc i} would be much stronger and 
C{\sc iv} line will not be present. The Br$\gamma$ and He{\sc ii}
2.188\,$\mu$m lines are almost equally strong. The extinction corrected
$K_S$ magnitudes of the stars are 9.9 and 10.04\,mag. Comparison with 
the smoothed to R=9000 \cite{han05} templates shows, that Obj1 and Obj2 are 
similar to CygOB2$-$8C-O5If and HD46150-O5Vf. Unfortunately, the S/N is not very 
high and for the most part in many spectral ranges, dwarfs, giants, and many 
supergiants look nearly the same.  Our classification for Obj1 and Obj2 is 
early O5-O6 spectral type, but  it is hard to distinguish between the 
luminosity class I and V.  

 Object\,3 and Object\,5 show He{\sc ii} 2.188\,$\mu$m stronger than
than in Obj1 and Obj2, and therefore, higher temperature. The stars are
similar to HD66811-O4If and HD46223-O4Vf. Obj3 is slightly
brighter than Obj1 and Obj2, while Obj5 is fainter. Our spectral 
classification is O4-O5, luminosity class I or V.  

 Object\,4 shows Br$\gamma$ and He{\sc i} lines in strong emission, the He{\sc ii} line is at the very end of our setup and there is no C{\sc iv} emission.  Based on this spectrum, it is extremely  hard to tell its spectral type. The star is saturated in our H and $K_S$ images and have no 2\,MASS measurements because of crowding. We can estimate approximately its $K_S$ magnitude, using narrow-band Br$\gamma$
and Spitzer 3.6 mag values (see Table\,1). The mean $K_S$ from these estimations
is 9.2 mag, which corrected for the mean extinction derived from Table~\ref{reddening} gives $K_S$$_0$=7.4. That is a magnitude and half brighter than Obj 8, which is classified as 'evolved' O4-O5 If supergiant (see below). So, Obj 4 is even more evolved from the main sequence and thus probably more massive (assuming co-eval star formation) than Obj 8. If what we see on the red part of the spectrum is really HeII 2.189, then the star is similar to WN7 objects (Fig. 4 of Figer et al. 1997, WR131). If, on the contrary, the strong line at 2.189 is artificial, then the star could be a Ofpe/WR9 like HDE 269582 from the Morris et al. (1996) atlas and to IRS34W (Martins et al. 2007, Fig. 2) or  Fig. 3 of Crowther \& Bohannan (1997). In addition, the wind velocity is low ($V_\infty \sim$ 350 km/s), as suggested  by the HeI 2.112 $\mu$m line, which also supports the Ofpe classification.  Even though more observations are necessary to clarify the nature of this interesting object, which is the brightest in our sample. Our tentative classification for Obj4 is Ofpe/WN9.

 Object\,6 is probably a massive Young Stellar Object (YSO). The Br$\gamma$ is in
strong emission, Mg{\sc ii} at 2.138 and 2.144\,$\mu$m can
be identified. This is the faintest object in the sample and,
according to the color-color diagram, shows IR excess.

  Object\,8 shows Br$\gamma$ in emission and also strong
He{\sc ii} 2.188\,$\mu$m in absorption. 
The star is spectroscopically similar to HD15570, 
classified as O4If \cite{han05}. It is significantly
brighter than Obj1, 2, 3, and 5 (the extinction corrected $K_S$=9.10 mag).
For these reasons, we believe the star may be a supergiant.

\subsubsection{Absorption line stars}

\begin{figure}
\resizebox{\hsize}{!}{\includegraphics{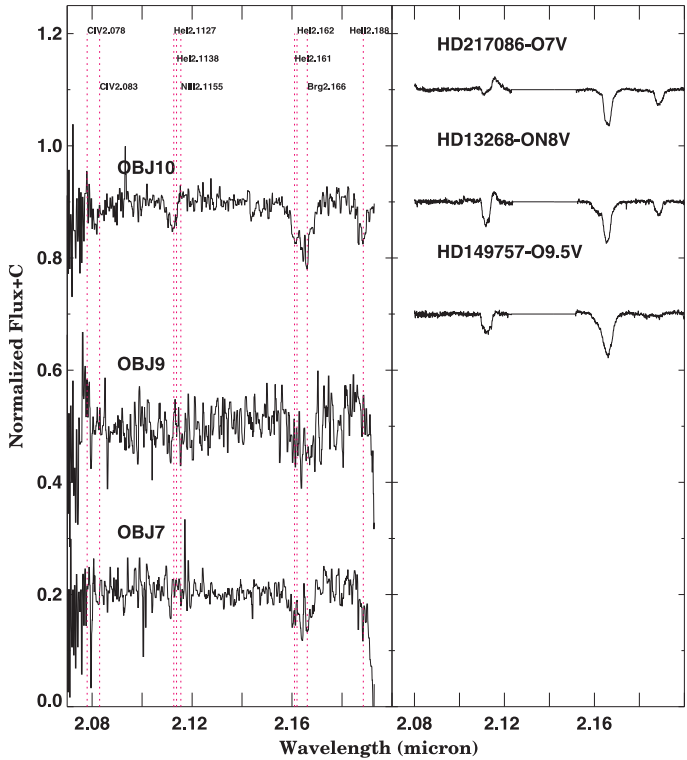}}
\caption{Left panel: The absorption line stars of [DBS2003]\,179.
Right panel: The template spectra taken from Hanson et al. (2005).
The symbols are the same as in Fig.~\ref{db179_sp_em}.
}
\label{db179_sp_abs}
\end{figure}

The spectra of the absorption line stars are shown in Fig.~\ref{db179_sp_abs}.
Object\,10 shows He{\sc i} 2.1112\,$\mu$m, He{\sc ii} 2.188\,$\mu$m
and Br$\gamma$, N{\sc iii}/O{\sc iii} might exist weakly and  
the He{\sc ii} line is relatively strong. The star
is similar to HD13268$-$ON8V type. The extinction corrected $K_S$ 
mag is 10.79 mag. This allows us to put the spectral type of 
Obj10 as O8 V.

It is impossible to classify the stars Obj7 and Obj9. They have
noisy spectra (S/N=25) and the only visible feature is Br$\gamma$
in absorption. Note that from the radial velocity analysis (see below)
these stars are unlikely to be cluster members. 

\begin{table*}\small
\begin{center}
\caption{Spectral classification of the observed stars in
[DBS2003]\,179.}
\label{stars_spectra}
\begin{tabular}{lrrccccrc}
\hline
& & & & & & & & \\[-6pt]
\multicolumn{1}{l}{Object} &
\multicolumn{1}{c}{S/N} &
\multicolumn{1}{c}{C{\sc iv}} &
\multicolumn{1}{c}{N{\sc iii}/O{\sc iii}+He{\sc i}} &
\multicolumn{1}{c}{HI(Br$\gamma$)} &
\multicolumn{1}{c}{He{\sc ii}} &
\multicolumn{1}{c}{Spec.type} &
\multicolumn{1}{c}{V$_{\sc rad}$} &
\multicolumn{1}{c}{$\sigma$(V$_{\sc rad}$)}\\
\multicolumn{1}{l}{} &
\multicolumn{1}{c}{} &
\multicolumn{1}{c}{2.078$\mu$m} &
\multicolumn{1}{c}{2.116$\mu$m} &
\multicolumn{1}{c}{2.661$\mu$m} &
\multicolumn{1}{c}{2.185$\mu$m} &
\multicolumn{1}{c}{K-band} &
\multicolumn{1}{c}{km\,s$^{-1}$} &
\multicolumn{1}{c}{km\,s$^{-1}$}\\
& & & & & & & & \\[-8pt]
\hline
& & & & & & & & \\[-6pt]
obj1 & 99&Em&$-$2.8&    0.8&2.1&   O5-O6I-V& $-$96& 12\\
obj2 & 75&Em&$-$2.3&    3.7&2.3&   O5-O6I-V& $-$75& 13\\
obj3 & 53&Em&$-$2.9&    3.8&2.3&   O4-O5I-V& $-$98& 12\\
obj4 & 84&Em&$-$6.0&$-$37.2&\multicolumn{1}{c}{--}  &   Ofpe/WN9&$-$163& 12\\
obj5 & 74&Em&$-$2.9&    3.4&3.2&   O4-O5I-V& $-$54& 14\\
obj6 & 25&NP&    NP&$-$12.1& NP&       YSO& $-$79&18\\
obj7 & 50&NP&    NP&    5.7& NP&        --&$-$214& 14\\
obj8 &112&Em&$-$2.9& $-$1.4&1.8&   O4If-O5If& $-$74& 14\\
obj9 & 25&NP&    NP&    6.0& NP&        --&   258&20\\
obj10& 79&NP&   2.4&    7.9&2.0&   O8V    & $-$60& 12\\
& & & & & & & & \\[-8pt]
\hline
\end{tabular}
\end{center}
\end{table*}

\subsection{Radial velocities}

\begin{figure}
\resizebox{\hsize}{!}{\includegraphics{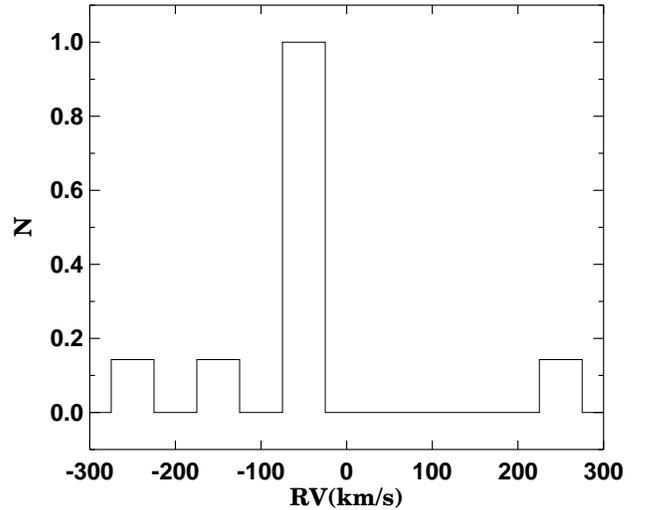}}
\caption{The radial velocity histogram.}
\label{db179_rvhist}
\end{figure}

We derived the radial velocities of all ten stars with the
IRAF task {\tt fxcor}, which uses a cross-correlation Fourier
method. Since the stars show very different characteristics,
several radial velocity standards are used: for the stars Obj1,
Obj2, Obj3 and Obj5 we used HD15629, an OV5 spectral type star
with V$_{\sc rad}$=$-$48\,km\,s$^{-1}$; Object\,8 was
cross-correlated with HD15570, an O4If spectral type star with
V$_{\sc rad}$=$-$15\,km\,s$^{-1}$; Obj6 and Obj 4 are measured
by comparison with DM+493718, an Be spectral type star; all absorption 
line stars are compared with HD191639 (B1V, V$_{\sc rad}$=$-$7\,km\,s$^{-1}$)
and HD167785(B2V, V$_{\sc rad}$=$-$10.6\,km\,s$^{-1}$). The
spectra are taken from Hanson et al (2005) and \cite{han96}
spectral atlases. The HD15629 have been shifted to zero velocity,
as measured in air using the sophisticated atmospheric modeling 
(Repolust et al. 2005). In this study, the high resolution spectra of 
Hanson et al. (2005)
were modeled to derive directly absolute values of luminosity and 
temperature. The models (synthesized spectra) are made at zero 
velocity and thus the modelers shift the observed spectrum to align 
them with the models. It is accurate to $\sim$10\,km\,s$^{-1}$. 
The radial velocities and their estimated errors are given in cols.~8 and
9 of Table~\ref{stars_spectra}. 
The membership of the stars is determined on the basis of the radial velocity 
histogram (see Fig.~\ref{db179_rvhist}). Most of the stars are concentrated 
within the radial velocity interval ($-$100,\,$-$50) 
km\,s$^{-1}$.  The stars Obj7 and Obj9 have very different radial velocities and
most probably are back- or foreground stars. The photometry also confirm the non-membership
of these stars: the extinction corrected $K_S$ magnitudes and $J-K_S$ colors are (11.05; -0.3) 
and (7.06,1.29)\,mag (see Tables\,1 and \,2)
and they are situated far from the determined cluster boundaries.
The Object\,4 also has a relatively high radial
velocity; the reason could be the lack of a suitable template 
for comparison or the peculiar nature of the star.
 The mean radial
velocity value is then calculated from seven
cluster members (Obj4 is not taken in consideration) and is
V$_{\sc rad}$=$-$77$\pm$6 ~\,km\,s$^{-1}$.

\section{Physical Parameters of [DBS2003]\,179}

\begin{figure}
\resizebox{\hsize}{!}{\includegraphics{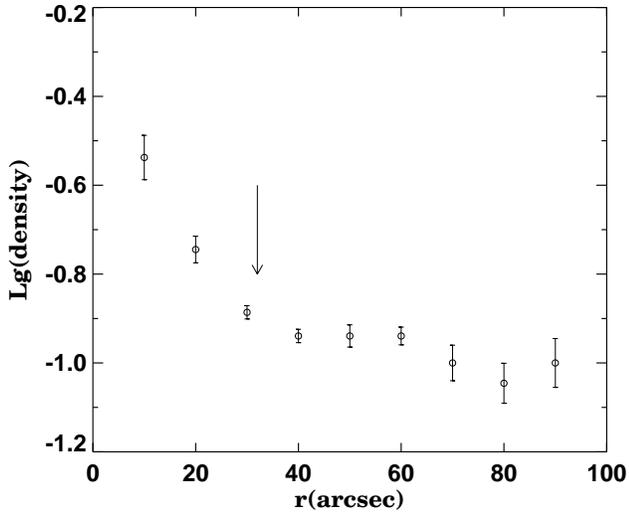}}
\caption{Radial surface density (in number of stars per square
arcsec) profile of [DBS2003]\,179. The bars show the 3$\sigma$
Poison uncertainties. The arrow indicates the derived
32\,arcsec cluster radius.}
\label{db179_radius}
\end{figure}

The cluster seems to be compact and dense. To determine its
boundaries we performed direct star counting, assuming spherical 
symmetry.  The projected star
number density as a function of radius is shown in Fig.~\ref{db179_radius}. The cluster
boundary was determined as the radius at which the density
profile exceeds twice the standard deviation of the surface
density in the surrounding field. This yields a radius of
32\,arcsec. 

With the newly-adopted cluster radius, we statistically decontaminated the
``cluster''+``field'' color-magnitude diagram by removing as many stars as 
are present on the ``field'' color-magnitude diagram. The details of 
the procedure are described in  Borissova et al. (2005). We have selected 180 
potential cluster members.

To determine the distance to the cluster, we used the K-band spectral
classification of our objects and the intrinsic color and absolute magnitudes 
of the main sequence and supergiants stars 
given in the latest calibrations of Martins \& Plez (2006). 
We feel most confident about the spectral type and luminosity class of
Obj10, as an O8V.  Likewise, we feel moderately confident about the classification
of the O supergiant, Obj8, showing weak emission in Br$\gamma$.
However, we do not share the same confidence in knowing the luminosity 
class for Obj1, 2, 3, and 5.  In Table 4 we have calculated
the distance moduli for each object assuming two luminosity classes for
Obj1, 2, 3, and 5 - as supergiants (I) and dwarfs (V). 
As can be seen from Table~\ref{distance}
if we assume the luminosity class I for these four objects, the distance 
modulus is more than one magnitude larger than this value calculated
from Obj8 and Obj10. All four of these stars have $K_0$ magnitudes 
of about 10 and the fact that these stars show both Br$\gamma$ 
and HeII in absorption, it seems likely they are main sequence stars.  
Finally, the star showing the most extreme spectrum, Obj4, also has the largest
K$_{s0}$ magnitude, estimated to be around 8.  This object is 
likely an extreme supergiant (Of+).
These luminosity designations of dwarf, supergiant, and Of+ are consistent
with their spectra and the individual distance moduli they predict. 
Using them, we adopted the average ($m$$-$$M$)$_0$=14.5$\pm0.3$\,mag or 
7.9 Kpc. The error is calculated as a standard
deviation of the mean value and quadratically added errors from
photometry and the variance due to unsure spectral classification.
This is a larger
distance than the value of 5\,Kpc estimated in
Paper~I.  The reason for the difference is in the a-priori
assumption used in the 10$^{\rm th}$ star method for the spectral type
of the 10$^{\rm th}$ brightest cluster member.

\begin{table*}
\begin{center}
\caption{Distance to the cluster.}
\label{distance}
\begin{tabular}{rcccccc}
\hline
& & & & & & \\[-6pt]
\multicolumn{1}{r}{Object} &
\multicolumn{1}{l}{$K_0$} &
\multicolumn{1}{l}{Spec. type} &
\multicolumn{1}{l}{Lum. class} &
\multicolumn{1}{l}{$(m-M)_0$} &
\multicolumn{1}{l}{Lum. class} &
\multicolumn{1}{l}{$(m-M)_0$} \\
& & & & & & \\[-8pt]
\hline
& & & & & & \\[-6pt]
obj1 &9.90&O5&I&15.44&V&14.31\\
obj2 &10.04&O5&I&15.45&V&14.48\\
obj3 &9.79&O4&I&15.31&V& 14.48\\
obj5 &10.17&O5&I&15.71&V&14.59\\
obj8 &9.10&O4&I&14.65&--& --\\
obj10&10.79&O8&--&--&V&14.41\\
& & & & & & \\[-8pt]
\hline
\end{tabular}
\end{center}
\end{table*}

\begin{figure}
\resizebox{\hsize}{!}{\includegraphics{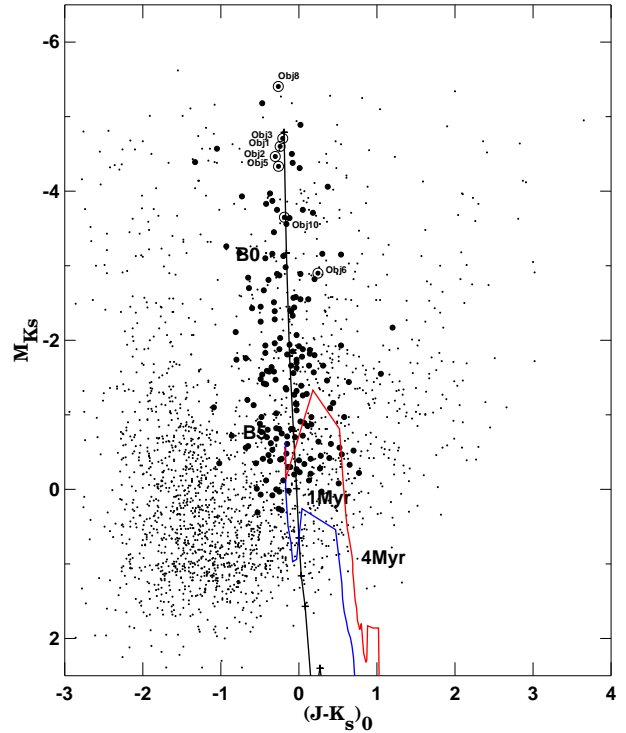}}
\caption{The $M_{K_S}$ versus ($J$$-$$K_S$)$_0$ color-magnitude
diagram of [DBS2003]\,179. All stars in the field of view are
shown as small circles, while the cluster members are plotted
with large filled circles. The unreddened main sequence
(\cite{sch82}) is drawn with a solid line. The adopted distance
modulus is ($m$$-$$M$)$_0$=14.5\,mag. The stars with know
spectral types are labeled and are corrected with their
individual reddening values. The 1 and 4\,Myr PMS isochrones are taken
from Siess et al. (2000).
}
\label{db179_cmd_cl}
\end{figure}

The $M_{K_S}$ versus
($J$$-$$K_S$)$_0$ color-magnitude diagram of the cluster is
plotted in Fig.~\ref{db179_cmd_cl}. We adopted E($B$$-$$V$)=5.2 and
($m$$-$$M$)$_0$=14.5$\pm$0.3\,mag.
The unreddened main sequence (\cite{sch82}) is shown with the
solid line as well as the isochrones of pre-main sequence stars
(PMS) of 1 and 4\,Myr (Siess et al. 2000). The stars with known
spectral type are plotted with large circles and labeled. In
general, as stated above, their absolute $K_S$ magnitudes are 
consistent with luminosity class V (Obj1, Obj2, Obj3, Obj5, and Obj10) 
and I (Obj8).

Our photometry is not deep enough to properly sample the 
pre-main-sequence stars in [DBS2003]\,179 and derive an age based on isochrones
shown in Fig.\ 9.  However, the presence of O type supergiants and 
the absence (so far discovered) of WC stars and/or red supergiants 
implies an upper limit to the age of 4-5\,Myr (Maeder \& Meynet 2003).  
As a lower age limit we adopt 2\,Myr taking into account the presence 
of at least one Ofpe/WN9 type star (Obj4) (Crowther \& Dessart 1998; Figer et al. 1999, 2002;
Blum et al. 2001; Martins et al. 2007). For comparison we also plotted 
(Fig.~\ref{db179_cmd_cl}) the  pre-main sequence isochrones from
Siess et al. (2000). Thus, the present stellar population of [DBS2003]\,179 
appears consistent with an age of order 2-5\,Myr.
 
We determine the total stellar mass of the cluster in the following 
way. First, assuming the Salpeter initial mass function, masses of O 
supergiants 50\,$M_{\odot}$ 
(Herrero et al. 2002, Hillier et al. 2003) and masses of 
O-type dwarfs 30\,$M_{\odot}$ (Bouret et al. 2003),
we calculated a total mass of 250\,$M_{\odot}$
for the spectroscopically-confirmed cluster members.
Second, using $A_V$=16.6, ($m$$-$$M$)$_0$=14.5\,mag, and integrating
along the 5\,Myr Padova isochrone (Girardi et al. 2002), we
obtain a total present-day mass of the observed main sequence cluster members
down to 12\,$M_{\odot}$ $\sim$3100\,$M_{\odot}$. 
Integrating over 2.5\,Myr isochrone gives for the same mass interval
$\sim$3500\,$M_{\odot}$. In summary, we will adopt the mass of all 
observed stars of $\sim$3550\,$M_{\odot}$. Since no completeness correction has 
been applied this value should be considered as lower mass limit of the cluster.
Our photometry is not deep enough and a lot of cluster members are probably not 
yet seen. Nevertheless, some estimation of the expected total cluster mass can be 
calculated
adopting the Salpeter law, normalizing it to the upper three mass bins 
(encompassing stars with masses about 10 solar masses), 
and integrating down to 0.8 solar masses. This 
yields of 0.7$\times$10$^4$\,$M_{\odot}$. The total number of stars
determined with this procedure is 1240.

\section{Summary and Conclusions}

To place [DBS2003]\,179 in the context of the currently known
Milky Way supermassive/starburst clusters, we compared it with
the Westerlund\,1/2, the Orion Nuclear Cluster (ONC), NGC3603, Arches, and 
R136 clusters. The comparison is given in Table~\ref{comparison}, 
reproduced from Table~9 in Brandner
et al. (2007). The table rows present: age, relaxation time
t$_{\rm relax}$, half-mass radius r$_{\rm hm}$, total number of
cluster members N$_{\rm tot}$, and total cluster mass M$_{\rm tot}$.

To calculate the relaxation time of the cluster,
we use the method described in Sect. 5.2 of Brandner et al. 
(2007). First, we calculated the velocity dispersion 
$\sigma$ as: 
$$\sigma = \sqrt{0.4 \times {\rm G} \times {\rm M_{cl}} / {\rm r_{hm}}} $$

The half-mass radius of the cluster is taken as the characteristic radius
(Portegies Zwart et al. 2002) and is estimated as 0.2 pc.
We derive a 1-D velocity dispersion $\sigma$=8.5$\,{\rm km/s}$. 
This is in good agreement with the velocity dispersion of 6 km/s 
derived from the radial velocities measurements in Sect. 3.
Then, we calculated a crossing time for a star located at the half-mass radius 
and with a velocity equal to the velocity dispersion  
t$_{\rm cross} = {\rm r_{hm}} / \sigma$ = $2.2 \times 10^5$\,yr.
Thus, the half-mass relaxation time
$$ {\rm t}_{\rm relax} = \frac{0.2 {\rm N_{\rm tot}}}{\ln 
(0.1 {\rm N_{\rm tot}})} \times \sqrt{0.4} \times {\rm t}_{\rm cross} = 7 \times 10^6\,{\rm yr},$$ 
where N$_{\rm tot}$ is a number of the cluster members.   


\begin{table*}[htb]
\begin{center}
\caption[]{Comparison of [DBS2003]\,179 with other young, massive
clusters. The compilation data are from Brandner et al. (2007;
Table~9). The footnotes list the sources for the data on each individual
cluster. 
 
}
\label{comparison}
\begin{tabular}{lccccccc}
\hline
& & & & & & \\[-6pt]
&Wd1$^a$ &ONC$^b$           &Wd2$^c$          &NGC3603YC$^d$       &Arches$^e$          &R136 cluster$^f$&[DBS2003]\,179          \\
& & & & & & \\[-8pt]
\hline
& & & & & & \\[-6pt]
Age [Myr]     &3.6$\pm$0.7    &0.3--1.0          &1--3             &2.5                &2--3                &2--3            &2.5--5                  \\
t$_{\rm relax}$ [Myr]     &400            &2.1               &                 &10-40                   &2?                  &                &7                       \\
r$_{\rm hm}$ [pc]         &1.0            &0.8               &                 &0.7-1.5          &0.24                &1.1             &0.2                     \\
N$_{\rm tot}$             &10$^5$         &$>$2800           &                 &1.6 10$^4$             &$>$3450             &                &$>$1240                 \\
M$_{\rm tot}$ [$M_\odot$] &5$\times$10$^4$&0.18$\times$10$^4$&0.7$\times$10$^4$&1--1.6$\times$10$^4$&$>$1.3$\times$10$^4$&10$\times$10$^4$&(0.7--0.8)$\times$10$^4$    \\
& & & & & & \\[-8pt]
\hline
\end{tabular}
\begin{list}{}{}
\item[$^{\mathrm{a}}$] Brandner et al. (\cite{bra07}).
\item[$^{\mathrm{b}}$] Hillenbrand \& Hartmann (\cite{hillenbrand98}).
\item[$^{\mathrm{c}}$] Ascenso et al. (\cite{ascenso07}).
\item[$^{\mathrm{d}}$] Harayama et al. (2007).
\item[$^{\mathrm{e}}$] Figer et al. (\cite{figer99}); Stolte et al. (\cite{stolte05}).
\item[$^{\mathrm{f}}$] Andresen et al. (2007); Brandl et al. (\cite{brandl96}).
\end{list}
\end{center}
\end{table*}

\begin{figure}
\resizebox{\hsize}{!}{\includegraphics{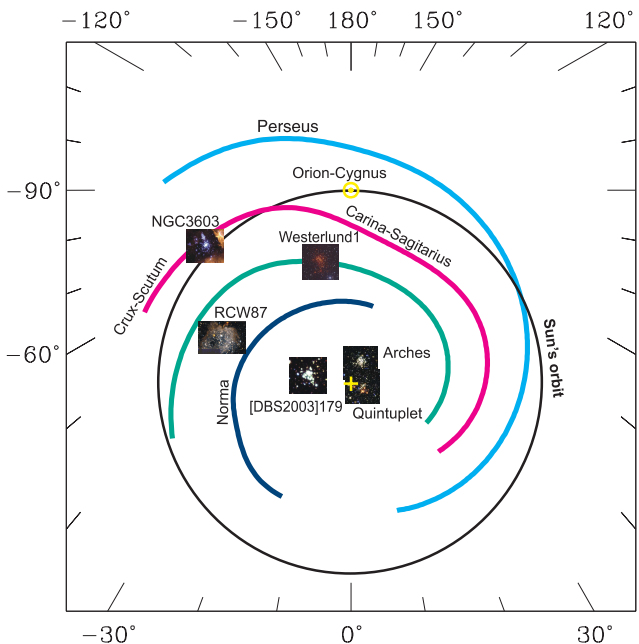}}
\caption{Location of the [DBS2003]\,179 and the other massive
Milky Way clusters.}
\label{db179_location}
\end{figure}

In summary, [DBS2003]\,179 is a young, not yet dynamically relaxed, dense and relatively massive cluster.  It has at least six O-type stars with masses in excess of 30\,$M_\odot$ and our rough estimate based on the most massive stars and assuming a Salpeter IMF suggests a cluster mass well in excess of 10$^3$, approaching 10$^4$\,$M_\odot$.  The location in the Milky Way of the [DBS2003]\,179 and the other clusters from Table 5 is shown in  Fig.~\ref{db179_location}.

Already, a number of optical and infrared clusters have been identified with characteristics similar to [DBS2003] 179, a few of which are listed in Table 5.  Presumably, all the optically observable clusters of this mass and age (any age) have been identified.  But how many more massive young clusters do we expect current near-infrared searches to find?  If our Milky Way galaxy is like every other normal spiral galaxy, its global star formation rate would suggest nearly 100 young clusters similar to [DBS2003]\,179 should currently exist within it (Weidner et al. 2005).  Just how could so many massive young clusters exist without our seeing them?  The 2\,MASS survey, mostly because of the extinction limitations at the J- and to a lesser extent the H-bands, is not seeing very deeply into our Galaxy.  Typical infrared clusters discovered with 2\,MASS have distances of between 1 and 2 kpc (Carpenter 2004).  Newer, deeper and longer-wavelength surveys look more promising and suggest that they are sampling to several kpc from the Sun (Mercer et al. 2007).  However, it is frequently overlooked that most of the star formation in our galaxy occurs deep within the molecular ring at a galactocentric distance of 4-5 kpc. Except at the closest tangent point to this structure, this constitutes a solar distance of more than a few kpc.  The deeply buried, very young massive cluster system of W49 may, in fact, exist within this ring.  However, no other known massive infrared (least optical) cluster is positively identified with this region of the Galaxy.  Surely, we are missing a significant number of massive clusters associated with this galactic structure.

Infrared searches must continue, but care must be given not to interpret the initial results too quickly.  Enormous biases in our current search methods are, in fact, leading to a highly skewed view of the massive cluster population presently in the Milky Way.  Until these biases are addressed, it will be difficult to fully appreciate the massive cluster population in our galaxy and relate it properly to the same phenomenon being observed in extragalactic super star cluster work. 

\begin{acknowledgements}
JB acknowledges support from Proyecto Fondecyt Regular \# 1080086.
D.M., J.B. and D.G. are supported by {\sl Centro de Astrof\'\i sica} FONDAP No. 15010003
and the Chilean Centro de Excelencia en Astrof\'\i sica
y Tecnolog\'\i as Afines (CATA). Support for RK is provided by Proyecto Fondecyt Regular 
\# 1080154 and DIPUV grant No 36/2006,  Universidad de Valparaiso, Chile. 
JB acknowledges support from DIPUV grant No 6/2007, Universidad de Valparaiso, Chile. 
MMH acknowledges support by the 
National Science Foundation under grant 0607497 to the University of Cincinnati. 
We are grateful for helpful comments from the referee.

\end{acknowledgements}

\end{document}